\documentclass[journal]{IEEEtran}
\usepackage{graphicx}
\usepackage{xfrac}
\begin{document}
\title{Ultra-precision DC source for \\
Superconducting Quantum Computer}
%
%

\author{Futian~Liang,
        Peng~Miao,
        Jin~Lin,
        Yu~Xu,
        Cheng~Guo,
        Lihua~Sun,
        ShengKai~Liao,
        Ge~Jin,
        and~ChengZhi~Peng 
\thanks{Manuscript received June 6, 2018.}
\thanks{Futian~Liang, Jin~Lin, Yu~Xu, Cheng~Guo, Lihua~Sun, ShengKai~Liao, and~ChengZhi~Peng are with the
Hefei National Laboratory for Physical Sciences at the Microscale and Department of Modern Physics, University of Science and Technology of China, Hefei 230026, China,
and Chinese Academy of Sciences (CAS) Center for Excellence and Synergetic Innovation Center in Quantum Information and Quantum Physics, University of Science and Technology of China, Shanghai 201315, China.}%
\thanks{Peng~Miao and Ge~Jin are with the
State Key Laboratory of Particle Detection and Electronics, University of Science and Technology of China, Hefei, Anhui 230026, P.R. of China.}%
\thanks{First author: Futian Liang(email: ftliang@ustc.edu.cn), Corresponding author: ShengKai Liao(email: skliao@ustc.edu.cn).}%
}

\maketitle
\thispagestyle{empty}

\begin{abstract}
The Superconducting Quantum Computing (SQC) is one of the most promising quantum computing techniques.
The SQC requires precise control and acquisition to operate the superconducting qubits.
The ultra-precision DC source is used to provide a DC bias for the qubit to work at its operation point.
With the development of the multi-qubit processor, to use the commercial precise DC source device is impossible for its large volume occupation.
We present our ultra-precision DC source which is designed for SQC experiments in this paper.
The DC source contains 12 channels in 1U 19~inch crate.
The performances of our DC source strongly beat the commercial devices.
The output rang is -7~V to +7~V with 20~mA maximum output current.
The Vpp of the output noise is 3~uV, and the standard deviation is 0.497~uV.
The temperature coefficient is less than 1~ppm/$^{\circ}$C in 14~V range.
The primary results show that the total drift of the output within 48h at an A/C room temperature  environment is 40~uV which equal to 2.9~ppm/48h.
We are still trying to optimize the channel density and long-term drift / stability.
\end{abstract}


\section{Introduction}
%
%
%
%
\IEEEPARstart{T}{he} quantum computing is well developed in the last decade for its great parallel acceleration characteristics.
The Superconducting Quantum Computing (SQC) is one of the most promising quantum computing techniques for the semiconductor technology-based manufacturing technique.
The SQC requires precise control and acquisition to operate the superconducting qubits~\cite{SQC_1,SQC_2},
such as
the fast and precise Arbitrary Waveform Generator (AWG) for control,
the fast and precise Data Acquisition (DAQ) for readout,
and the precision DC source for providing a DC bias for the qubit to work at its operation point.

The commercial precision DC sources are used at beginning in single or few qubits experiments.
With the development of the multiple qubits processor, to use the same DC source device is impossible for its large volume occupation.
The commercial devices are usually take large volume, such as only one output channel in a 3U rack size.
If we deploy those DC sources to Google's 72~qubit processor for example, only the DC sources could take a room.
There are no place to setup other devices, and the signal lines are too long to provide a good performance.
At beginning, there is no commercial precision DC source designed for SQC experiments,
and the commercial DC source are usually for power supply purpose and the voltage / current are more than sufficient for SQC experiments and accuracy / stability,
such as 1~ppm precision and $\pm$~5~ppm/24hour, $\pm$~35~ppm/year, are just meet the demands.
To optimize the large commercial device into a small size in a short time is also impossible.
So we design a SQC specified ultra-precision DC source.

\section{Design}

We study the requirements from the experiments with the SQC scientists,
and set a goal of 2~ppm precision and 10~ppm over all drift in -7~V to +7~V range to satisfy the demands of the experiments for our self-design ultra-precise DC source.
The DC source should archive the goal at least 24 hours against room temperature changes and other influence factors.

\subsection{Circuit Design}

The 2~ppm precision requires at least 19 bits vertical resolution.
The 2~ppm in 14~V range is 26.7~uV, and the output overall noise should slightly smaller than that.
To simple the design and avoid unnecessary influence, we use one single DAC instead of two master-slave DAC scheme \cite{2DAC}.
There are not too many choices for DACs with more than 19 bits.
We use the AD5791 from Analog Devices, Inc., and strictly following the  application note \cite{20-Bit-DAC} in our design.

 The voltage reference is also important for the design.
 To provides the precision and stability, only the LM399 and the LTZ1000 are the candidates.
 Still to simplify the design, we use LM399 for its simple setup components.
 We need a bipolar reference and to avoid the influence from the amplifier which is used in the polar conversion, we use two references to provide the bipolar reference.
 To avoid the unstable and "random jump" from the LM399 which we observed in the prototype design,
 we use three chips in parallel instead of one for one reference.

The ethernet is the best communication way for devices in SQC experiments.
The DC source requires very low data bandwidth for LAN communication,
and we choose an All-in-One ethernet solution,
the W7500P from the WIZnet Co., Ltd. which integrates an ARM Cortex-M0, 128KB Flash and hardwired TCP/IP core \& PHY for various embedded application platform.

\begin{figure}[!t]
\centering
\includegraphics[width=3.2in]{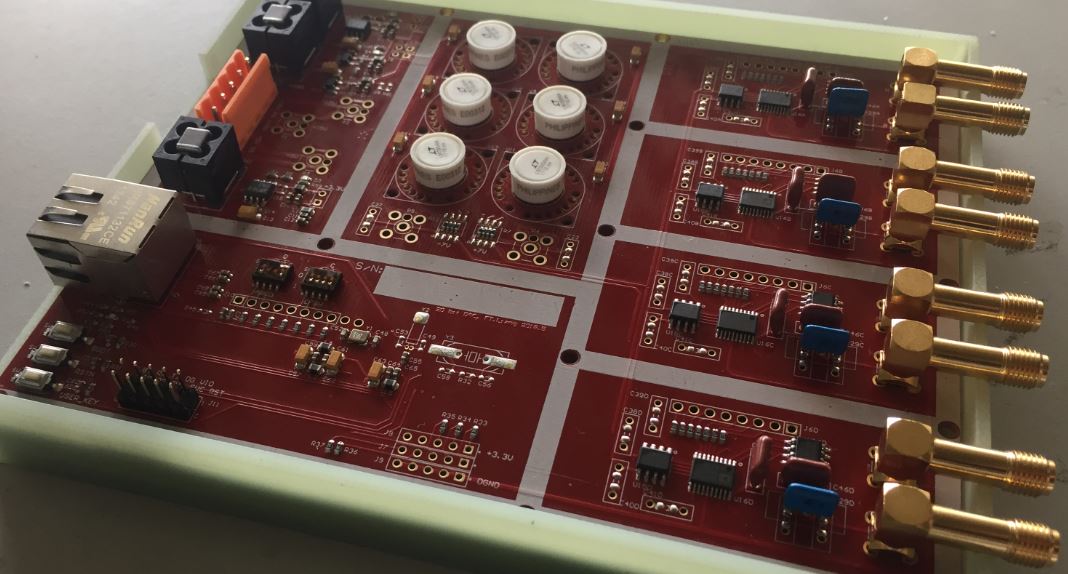}
\caption{The loaded PCB of the SQC specified ultra-precision DC source.}
\label{fig_pcb}
\end{figure}

In our prototype, we integrate four DAC channels, one group of reference and one ethernet into one unit, and three units together into one 1U 19~inch crate.
The loaded Printed Circuit Board (PCB) is shown in Fig.~\ref{fig_pcb}.

\subsection{Mechanical Structure Design}

For the prototype, we do not add temperature control in device.
However, we use several ways to stabilize the work surroundings of the circuits.

First, thermal isolation and windshields are used for the voltage reference.
A stabilizing heater is incorporated with the active Zener on a monolithic substrate which nearly eliminates changes in voltage with temperature in LM399.
The protections we deployed can reduce the influence from the PCB and airflow.

Second, a plastic shield is used for the whole PCB.
It provides the thermal isolation and windshield for components besides the voltage references.
The foamed plastic is insert into the gap between PCB and the shield.
The overall protection can smooth the temperature changes from the outside,
and provides a self-balance thermal environments for temperature sensitive components on PCB.
The above two ways make one DC source unit with a soft temperature change environment.

\begin{figure}[!b]
\centering
\includegraphics[width=3.5in]{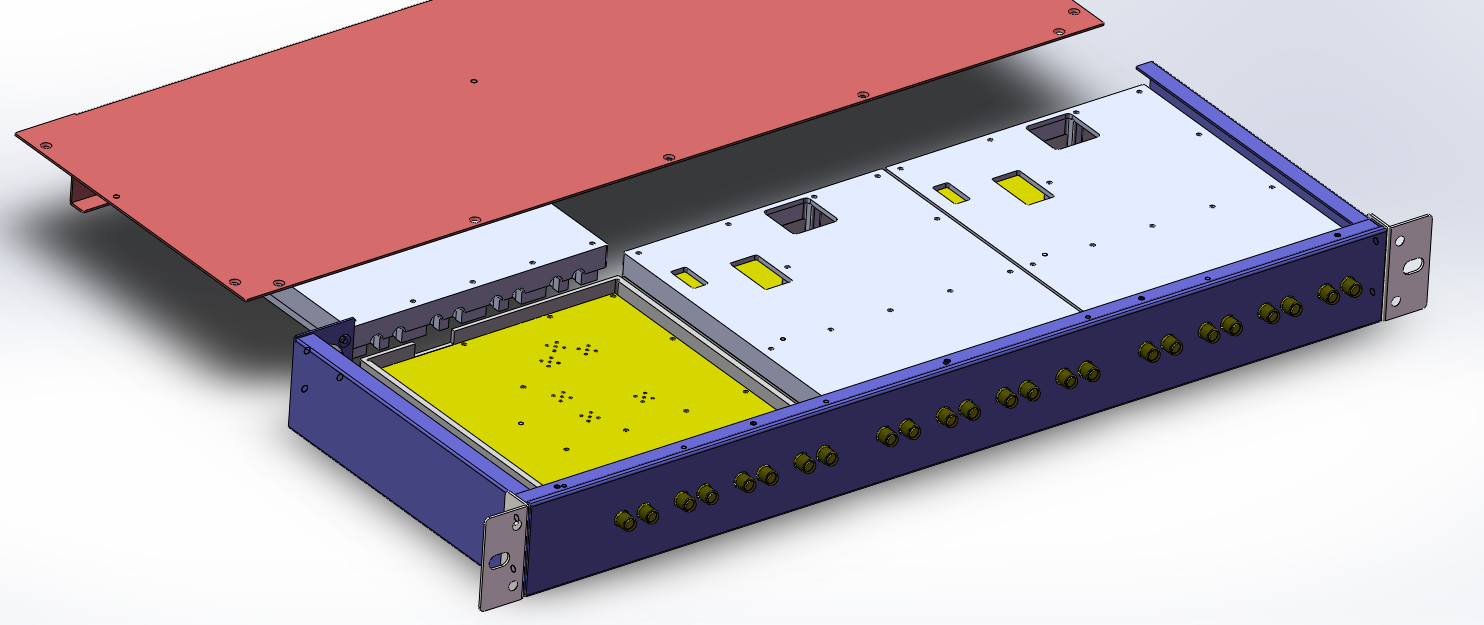}
\caption{Three DC source units with plastic shield and metal case.}
\label{fig_case}
\end{figure}

Finally, three units are covered with a metal case, shown in Fig.~\ref{fig_case}.
The metal case reduces the RF noise from the outside.

The design is compact,
and each unit contains four DC channels in a 13.2~cm(W) x 15.9~cm(L) x 3.5~cm(H) plastic shield.
Three units with 12 channels are packaged in a 1U 19~inch  crate (43.8~cm(W) x 16.3~cm(L) x 4.4~cm(H)) .

The power supply is from the external linear DC power source in the rack chamber.
+15~V and -15~V with less 600~mA each for analog circuits and +5~V with 200~mA for digital circuits.

\section{Results}

Our DC source is design for specified application use, not for technique study, and all results are from all effects in full device.

\subsection{Output Noise}

For precision DC source, low pass filter is usually deployed at the output, and the important noise is 0.1~Hz to 10~Hz.
In our design, we do not apply filters at the output for that there are low pass filters in the low temperature chamber in SQC experiments.
Even a passive RC filter can preform every well in the low temperature chamber which has only 10~mK  temperature variation.

\begin{figure}[!htb]
\centering
\includegraphics[width=3.5in]{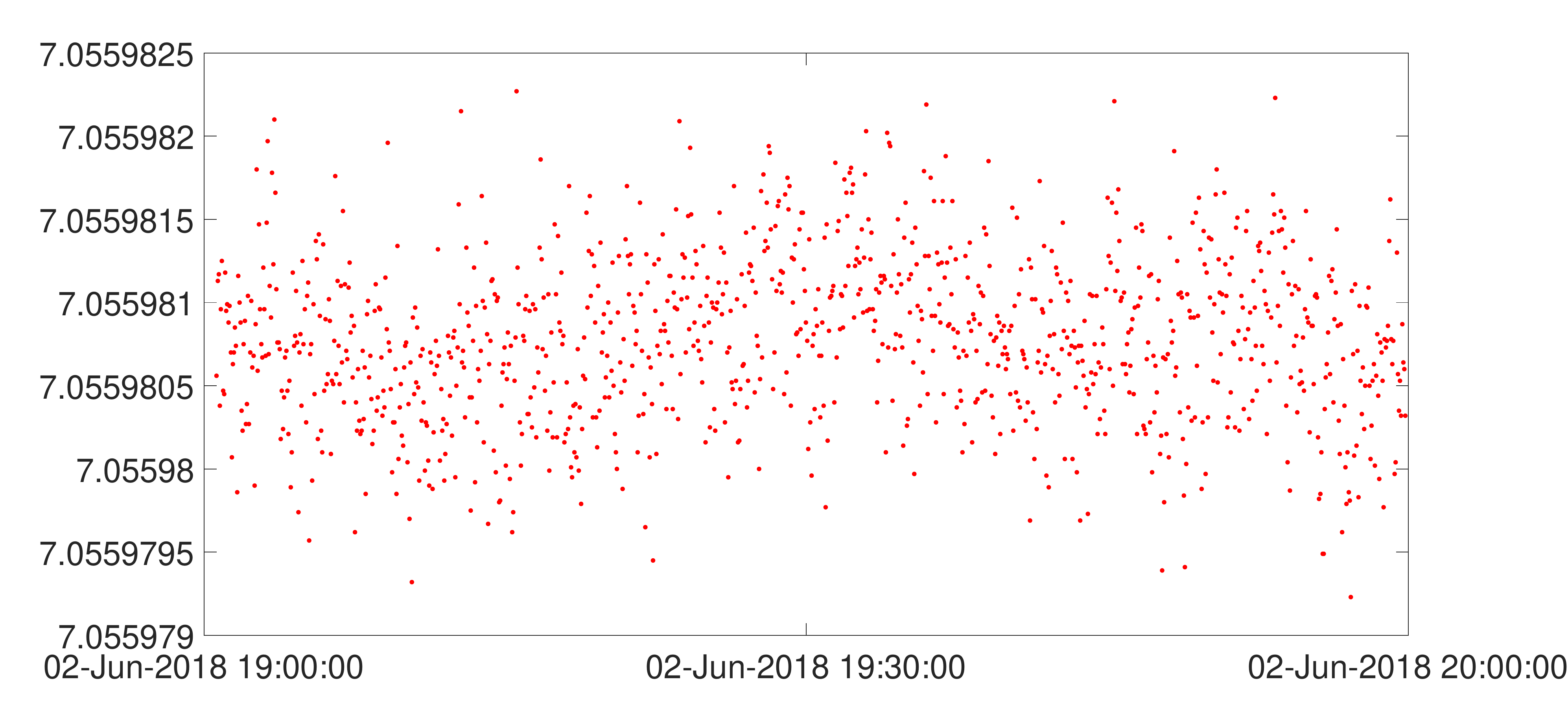}
\caption{+7~V output in one hour from Keysight 34470A. (7~\sfrac{1}{2} DMM)}
\label{fig_onehour}
\end{figure}

The output noise from the direct output of the DC source is the full bandwidth noise.
We use the integral time in voltage meter as the lowpass filter to limited the bandwidth in measurements.
With 10~V range and 10~NPLC in 7~\sfrac{1}{2} Digital Multimeter (DMM, Keysight 34470A), the resolution is 0.3~uV.
The one hour output values at +7~V are shown in Fig.~\ref{fig_onehour}, the Vpp is 3~uV which which equal to 0.21~ppm at 14~V range, and the standard deviation is 0.497~uV (0.004~ppm).

\subsection{Temperature Coefficient}

We did two kind of tests on temperature coefficient.

\begin{figure}[!t]
\centering
\includegraphics[width=3.5in]{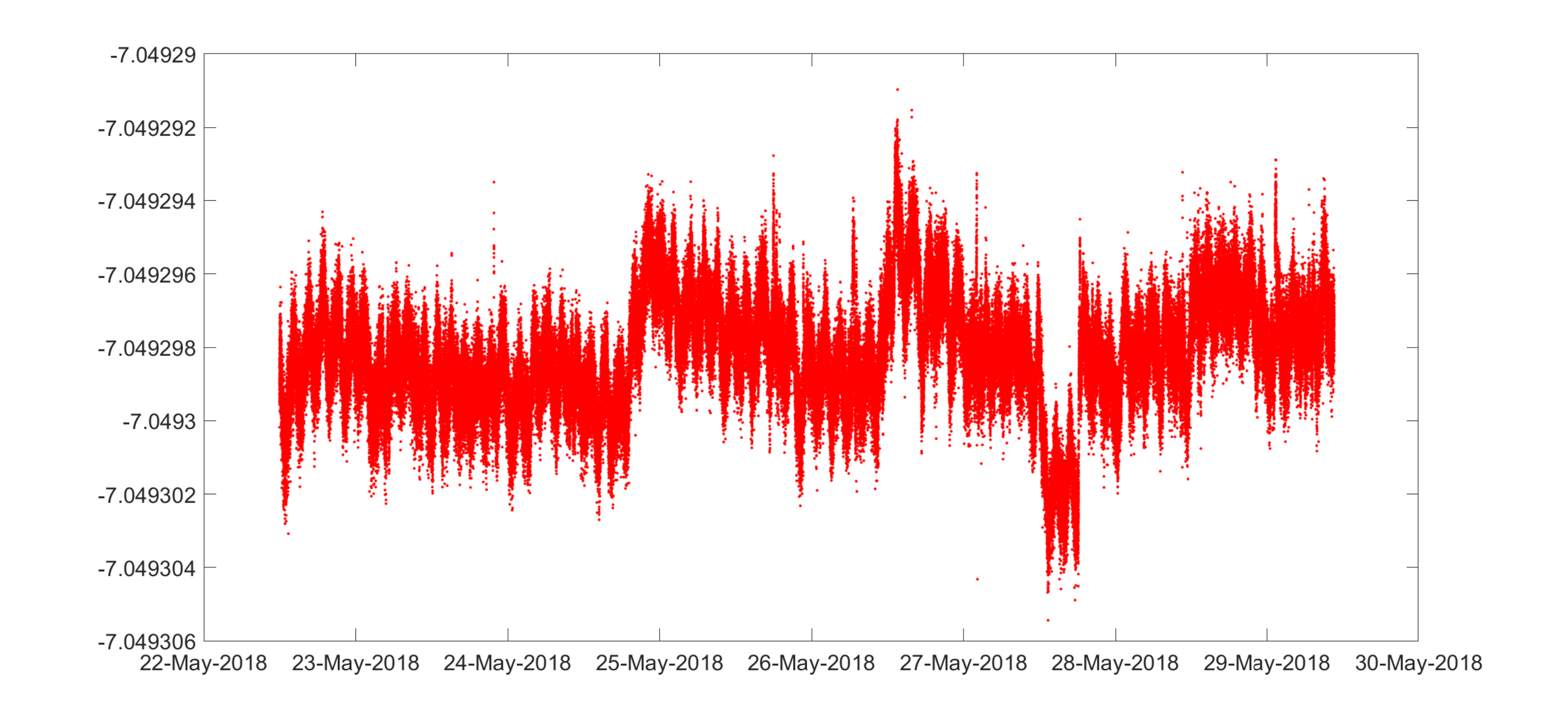}
a) Output variation in 1~$^{\circ}$C environment. (6~\sfrac{1}{2} DMM)
\includegraphics[width=3.5in]{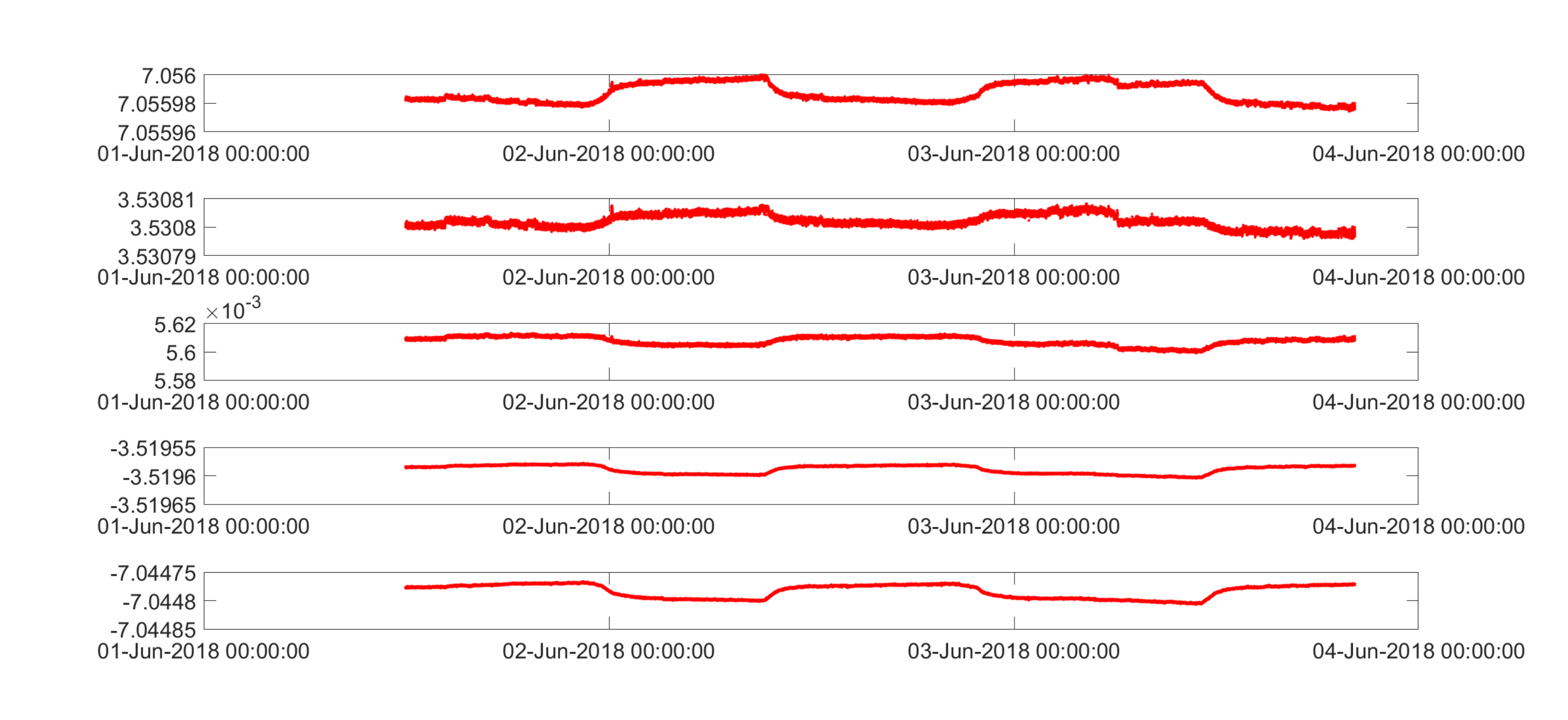}
b) Output variation in A/C room within 48~h. (7~\sfrac{1}{2} DMM)
\caption{Temperature Coefficient}
\label{fig_tc}
\end{figure}

A) In a strict temperature controlled room,
the room temperature is locked to 1~$^{\circ}$C variation.
The room temperature and output value at -7~V are shown in Fig.~\ref{fig_tc}~a.
The measured room temperature variation is 0.8~$^{\circ}$C peak-to-peak.
The RMS variation of -7~V in the same time is 11.5~uV (Vpp(14.5~uV) - nosie(3~uV)), which is equal to 14.4~uV/$^{\circ}$C, and 1~ppm/$^{\circ}$C.
When we zoom in the time range into one hour,
we can clearly see the temperature cycle is about 40~mins,
and the voltage output variation follows the cycle.
The temperature changes in this room are faster than we expect.
The laboratory room temperature or even a regular A/C room temperature change is more smooth than that.

B) In an A/C room, with a fix setting, the temperature variation is about 3~$^{\circ}$C or more over a day.
The outdoor temperature during the night is lower than A/C setting which enlarge the temperature variation indoor.
The outputs in 48~h are shown in Fig.~\ref{fig_tc}~b.
The worst overall drift Vpp is 40.6~uV at -7~V  which is equal to 2.9~ppm/48h in the A/C room.

\section{Analysis and Improvement}

Our DC source is already used in the SQC experiments.
The improvement of the experiment results will come out soon.
However, there are more work to do to optimize the design.

\subsection{Analysis}

It is the first time, we design the ultra precision design.
We can not take everything in the precision design into consideration，
such as the components are not well burn-in before use.
Even the equipment in measurements may be not perfect match the requirements.


The voltage reference in the 6~\sfrac{1}{2} DMM is the same precision and stable level as we use.
The measurements present both the stable and precision of the DMM and our design.
There are some "random" jump in our measurements results.
Only some of them are from our design.
We have to use two or more DMM to measure the same scanning-output at same time to distinguish the cause.

We try to use the 7~\sfrac{1}{2} DMM, and the Vpp of the output noise is reduced at least 0.5~uV the results from 6~\sfrac{1}{2} DMM.
The 7~\sfrac{1}{2} or 8~\sfrac{1}{2} DMM may help a lot during the test.

On qubit, the DC source is finally used as a current.
We use the same trace in SQC setup to measure the current,
and the precision is still the same.
If we can generate current directly, we may get an even better result.

\subsection{Improvement}

From the above design and test, the performance of our ultra-precision DC source well meets the requirements of the SQC experiments.
And the channel density is 30~times higher than the commercial devices on front panel, and 100~times in total volume occupation.

However, there is a long way to make our specified design into a general instrument. The followings are under our consideration:

1) Long-term drift.
We do not have time to evaluate the drift in super long time eg. a year in this paper.
However, the experiments are counts on the device more than years.
We should evaluate and calibrate the device termly.

2) Output filter.
The SQC experiments have their own lowpass filter in the low temperature chamber.
For our design, an on-site output filter can great lower the output noise, and help the output performance.
However, the stable and temperature coefficient of the filter at the expect accuracy is also challenge.

3) Consistency of all channels.
In our design, all components after the voltage reference, such as the amplifiers, the DACs, the resistors or even the PCB traces can involve the difference among channels.
Those make the output of each channel can not perfect match each other.
We should make efforts to reduce the mismatch of each channel.

4) Calibration.
The SQC experiments use the scan mode to find the best DC value,
and the exact value of each code is not necessary.
For general device, the exact value is useful.
We should do calibration and correct the output.

5) Temperature feedback.
It is the simple way to overcome the  temperature coefficient.
And it is not a easy way.

6) Output capability.
The output range now is limited by the voltage reference, and the current are just meet the SQC experiments.
We need do more if an integer voltage range, such as +10~V to -10~V, or more current driving is required.



\section{Conclusion}

To satisfy the demands of the ultra-precision DC source on channel density, precision and stability in SQC measurements,
we design our own ultra-precision DC source.
The DC source contains 12 channels in 1U 19 inch crate.
The performances beat the commercial devices, such as
the 20 mA maximum output current in -7~V to +7~V output rang,
the 0.497~uV standard deviation of the output noise,
the 1~ppm/$^{\circ}$C of the temperature correlation.
We are still trying to optimize the channel density and long-term drift and stability.

\section*{Acknowledgment}

We thank the QuantumCTek Co., Ltd.  for the support on mechanical structure design, loading PCB and fabrication.



\begin{thebibliography}{2}

\bibitem{SQC_1}
Chen, Yu, et al. \emph{Multiplexed dispersive readout of superconducting phase qubits.} Applied Physics Letters 101.18(2012):104401.

\bibitem{SQC_2}
Ofek, N, et al. \emph{Extending the lifetime of a quantum bit with error correction in superconducting circuits.} Nature 536.7617(2016):441-445.

\bibitem{2DAC}
Williams, Jim, et al. \emph{A standards lab grade 20-bit DAC with 0.1 ppm/C drift.} Linear Technology Corporation, Milpitas, Calif., Application Note 86 (2001).


\bibitem{20-Bit-DAC}
Egan, Maurice. \emph{The 20-bit DAC is the easiest part of a 1-ppm-accurate precision voltage source.} Analog Dialogue 44.4 (2010).


\end{thebibliography}
%

\end{document}